\documentclass[preprint]{aastex63}
\usepackage{amssymb,amsmath}
\usepackage{graphicx}
\usepackage{natbib}
\usepackage[OT1,T1]{fontenc}

\begin{document}

\title{Non-LTE Calculations of the \ion{Mg}{1} 12.32 $\mu$m Line in a Flaring Atmosphere}

\author{Jie Hong}
\affiliation{School of Astronomy and Space Science, Nanjing University, Nanjing 210023, People's Republic of China}
\affiliation{Key Laboratory of Solar Activity, National Astronomical Observatories, Chinese Academy of Sciences, 20 Datun Road, Beijing 100012, People's Republic of China}
\affiliation{Key Laboratory for Modern Astronomy and Astrophysics (Nanjing University), Ministry of Education, Nanjing 210023, People's Republic of China}

\author{Xianyong Bai}
\affiliation{Key Laboratory of Solar Activity, National Astronomical Observatories, Chinese Academy of Sciences, 20 Datun Road, Beijing 100012, People's Republic of China}
\affiliation{School of Astronomy and Space Science, University of Chinese Academy of Sciences, No. 19(A) Yuquan Road, Beijing 100049, People's Republic of China}

\author{Ying Li}
\affiliation{Key Laboratory of Dark Matter and Space Astronomy, Purple Mountain Observatory, Chinese Academy of Sciences, Nanjing 210033, People's Republic of China}

\author{M.~D. Ding}
\affiliation{School of Astronomy and Space Science, Nanjing University, Nanjing 210023, People's Republic of China}
\affiliation{Key Laboratory for Modern Astronomy and Astrophysics (Nanjing University), Ministry of Education, Nanjing 210023, People's Republic of China}

\author{Yuanyong Deng}
\affiliation{Key Laboratory of Solar Activity, National Astronomical Observatories, Chinese Academy of Sciences, 20 Datun Road, Beijing 100012, People's Republic of China}
\affiliation{School of Astronomy and Space Science, University of Chinese Academy of Sciences, No. 19(A) Yuquan Road, Beijing 100049, People's Republic of China}

\email{jiehong@nju.edu.cn}

\begin{abstract}
The infrared \ion{Mg}{1} lines near 12 microns are a pair of emission lines which are magnetically sensitive and have been used to measure solar magnetic fields. Here we calculate the response of the \ion{Mg}{1} 12.32 $\mu$m line during a flare and find that in our modeling this line has a complicated behavior. At the beginning of the flare heating, this line shows an intensity dimming at the line center. The intensity  then increases when heating continues, with increasing contributions from the heated layers in the chromosphere. The line formation height and the line width also increase as a result. As for the polarized line profiles, we find that flare heating tends to decrease the Zeeman splitting width and attenuates the Stokes $V$ lobe intensity. The wider features in the Stokes $V$ profiles are more pronounced during flare heating, which should be considered when performing magnetic field inversions. 
\end{abstract}

\keywords{}

\section{Introduction}
The infrared \ion{Mg}{1} emission lines at 12.32 and 12.22 $\mu$m (or 811.575 and 818.058 cm$^{-1}$) were first noticed by \cite{1981murcray} in high-resolution solar spectra. They were later identified as stemming from transitions between high Rydberg levels of the \ion{Mg}{1} atom \citep{1983chang}. Early observations \citep{1983brault} showed that these lines are visible in sunspot penumbrae or plages, while they disappear in sunspot umbrae. The line profiles typically show an emission peak and an absorption trough at the solar disk center, while the absorption trough disappears and the emission peak becomes stronger at the solar limb. These lines are also found to be very sensitive to solar magnetic fields, and can display a clear Zeeman splitting pattern in observations \citep{1983brault,1990deming,1993hewagama}. Thus, they can be a promising tool to diagnose magnetic fields, and there have already been many studies on this topic \citep{1990deming,1993hewagama,2000moran,2002jennings,2007moran}.

\cite{1992carlssonb} conducted a detailed study of the formation of these lines for the first time, and their results agreed well with the observations. They found that these lines are formed in the photosphere, and the emission feature is a consequence of population departure divergence of the high Rydberg levels. \cite{1995bruls} calculated the polarized line profiles, and found that there exist wide features in the Stokes $V$ profiles, corresponding to the absorption trough in the Stokes $I$ profiles. They also noticed that inverting the apparent magnetic field strength from $V$ splitting is more accurate than from $I$ splitting, especially for weak fields. 

Previous studies have suggested that major solar flares can have large influences on the photospheric lines \citep{2002ding,2018hong}. However, there are very few observations of a solar flare using the \ion{Mg}{1} 12.32 $\mu$m  line. \cite{1990deming} found that a flare can heat the umbra region so the line profiles in that region can also be in emission. They also found a brightness increase in the 12 $\mu$m continuum, while there seems to be no flare-related line broadening. \cite{2002jennings} found two distinct Zeeman splittings in the Stokes $V$ profile, and explained them as originating from opposite polarities in the flare trigger site. More observations are bound to be conducted with AIMS  \citep{2016deng} in the future.  Therefore, it is necessary to make theoretical calculations of the \ion{Mg}{1} lines under different conditions. As previous calculations of these lines are all based on the quiet-Sun atmosphere, it is required to perform new calculations based on a flare model.

In this paper, we perform non-local thermodynamics equilibrium calculations of the \ion{Mg}{1} 12.32 $\mu$m line in a flaring atmosphere. We introduce our method in Section~2, and present the results in Section~3. A discussion of the results is in Section~4, followed by a conclusion in Section~5.
 
\section{Method}
The 1D radiative hydrodynamics code RADYN \citep{1992carlsson,1995carlsson,1997carlsson,2002carlsson} has been widely used for flare simulations \citep{2015allred}. We use our previous flare models FHa and FHb in \cite{2019hong}, created with RADYN, to perform new calculations for the \ion{Mg}{1} 12.32 $\mu$m line. These two flare models characterize a small and an intermediate flare, which are supposed to be heated by a beam of non-thermal electrons with a hard energy spectrum. The beam heating rate rises linearly with time for 10 s, and the peak fluxes of the electron beams are 10$^{10}$ (model FHa) and 10$^{11}$ (model FHb) erg cm$^{-2}$ s$^{-1}$, respectively. 
Hereby we use the peak flux of the electron beam to refer to models FHa and FHb as models f10 and f11.

The magnesium atom is included in the background opacity in RADYN simulations, thus we use the RH code \citep{2001uitenbroek,2015pereira} to calculate the detailed radiative transfer of the \ion{Mg}{1} 12.32 $\mu$m line. We use the model atom \verb"MgI_66.atom" in the RH package that is adapted from \cite{1992carlssonb}, and update the values of the energy levels following \cite{1991kaufman}. The atmosphere in each RADYN snapshot is then fed to the RH code to calculate the desired spectral lines. As previously done \citep{2020hong}, we also take in the electron density and hydrogen populations from the RADYN simulation results to mitigate the default assumption of the statistical equilibrium for hydrogen and magnesium in RH. 

Apart from the calculation of the unpolarized line profiles, we also follow \cite{2018hong} to calculate the Stokes profiles when there is a magnetic field. The magnetic field is again assumed to be vertical and with an exponential distribution of $B(z)=B_0\textrm{e}^{-z/H}$. The values of $B_0$ and $H$ are specifically set so that the magnetic field is 2500 G at the height of 200 km, and attenuates to 1250 G at the height of 400 km. The magnetic field is static and does not change with time.

\section{Results}
\subsection{Model f10}
\begin{figure}
\epsscale{1.15}
\plotone{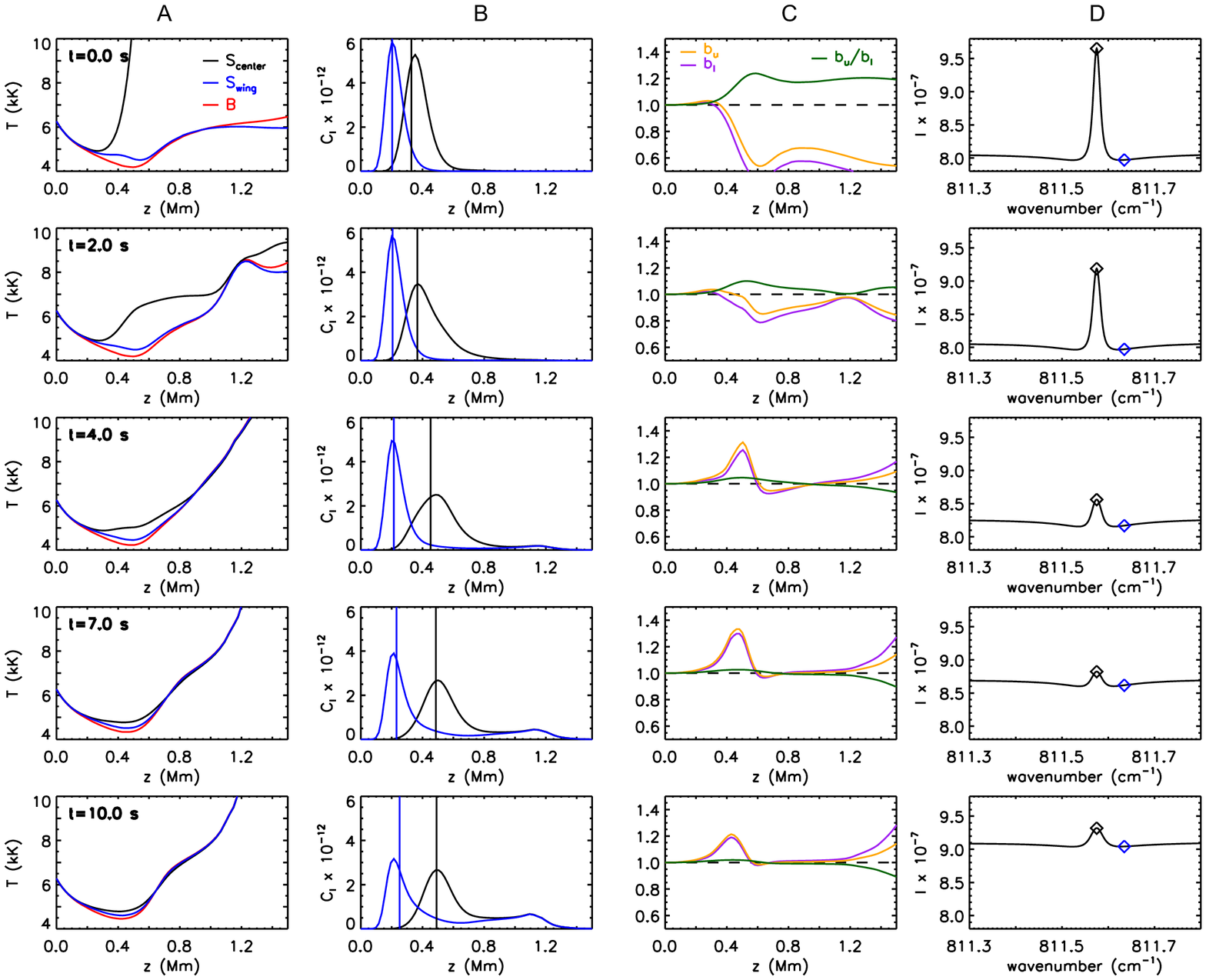}
\caption{The evolution of the atmospheric structure and line formation for the f10 model. Column A: Height distribution of the source function at the wavenumber points chosen in Column D (black and blue), and the Planck function (red), in the scale of excitation temperature. Column B: Height distribution of the contribution function $C_I$ at the same two wavenumber points. Vertical lines denote the height where the optical depth reaches unity. Column C: Height distribution of the departure coefficient of the upper level $b_u$ (orange) and that of the lower level $b_l$ (purple), and their ratio $b_u/b_l$ (green) for the \ion{Mg}{1} 12.32 $\mu$m line. Column D: The \ion{Mg}{1} 12.32 $\mu$m line profiles. The diamonds mark the wavenumber points that are chosen at the line center (811.575 cm$^{-1}$, black) and at the absorption trough in the blue wing (811.635 cm$^{-1}$, blue).}
\label{f10}
\end{figure}

We show the evolution of the atmosphere and line formation for the f10 model in Fig.~\ref{f10}. Before flare heating (see the top panels of Fig.~\ref{f10}), this line shows an emission peak and a wide absorption trough. We choose two wavenumber points, one at the line center (811.575 cm$^{-1}$), and the other at the absorption trough in the blue wing (811.635 cm$^{-1}$). The contribution function to the emergent intensity, defined as $C_I\equiv dI/dz$, shows that both the line center and the absorption trough are formed mainly in the photosphere. The chromosphere, above the temperature minimum region at about 500 km in our model, has a very small contribution to the intensity at the line center. The height where $\tau=1$ is about 360 km for the line center, and 200 km for the absorption trough. The formation height of the absorption trough is somehow similar to that of the \ion{Fe}{1} 6173 \AA\ line center \citep{2018hong}. The line source function is defined as 
\begin{equation}
S_\nu^l=\frac{2h\nu^3}{c^2}\frac{1}{(b_l/b_u)\mathrm{e}^{h\nu/kT}-1},
\end{equation}
 where the departure coefficients $b_l$ and $b_u$ denote how much the populations of the lower and upper levels deviate from the values under local thermodynamic equilibrium (LTE). One can see clearly that, at this time, the ratio of $b_u/b_l$ is larger than unity, so the line source function is also larger than the Planck function. Even if including the continuum, the total source function at the line center still has a large departure from the local Planck function, which contributes to the emission of the line. Our results of the \ion{Mg}{1} line before flare heating are in good agreement with the previous observations \citep{1983brault} and simulations \citep{1992carlssonb,1994rutten}.

\begin{figure}
\plotone{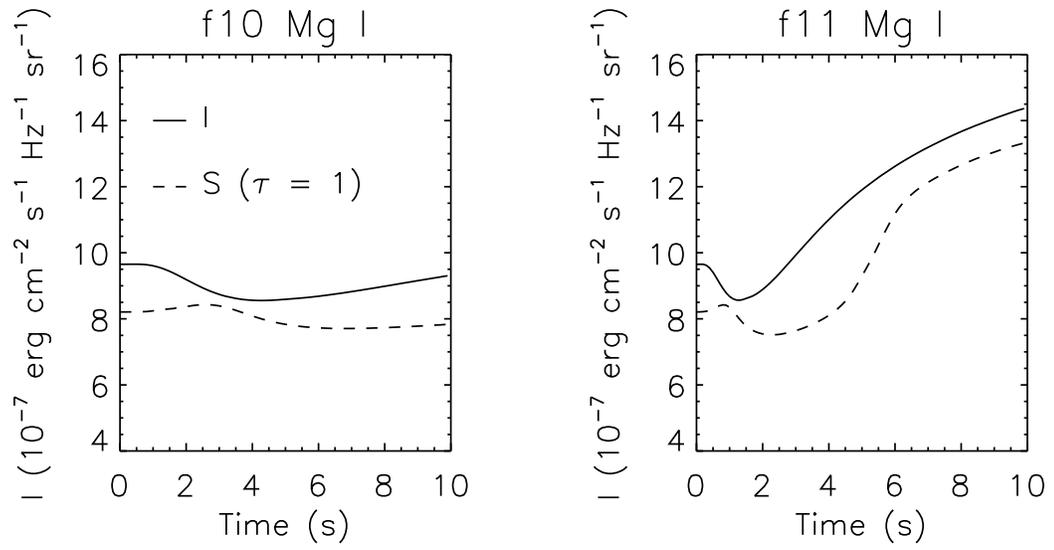}
\caption{Evolution of the line center intensity (solid) and the source function at the height of optical depth unity (dashed), for the \ion{Mg}{1} 12.32 $\mu$m line.}
\label{evol}
\end{figure}

From Fig.~\ref{f10} (Column D), one can see clearly that the intensity at the line center decreases at first and then increases. We show the evolution of the intensity at the line center in Fig.~\ref{evol}, and divide the total duration into two phases according to the change in the intensity at the line center. The first phase lasts from 0 s to 4 s in model f10. In this phase, the temperature of the chromosphere below 1.2 Mm increases rapidly as a result of flare heating (Fig.~\ref{f10}). The rise of the local electron density from flare heating increases the collisional recombination and  de-excitation rates. 
Under normal conditions $b_u>b_l$ because of the the replenishment flow resulting from overionization in combination with recombination/dexcitation through the high-lying Rydberg levels \citep{1992carlssonb,1994rutten}. When collisional (de-)excitation increases, $b_u$ and $b_l$ become almost equal, as the initial relative overpopulation in the upper level now has to compete with the collisional coupling between the upper and the lower state. This pulls the ratio of $b_u/b_l$ closer to one (essentially increased de-excitation corrects the non-LTE overpopulation from the recombination flow), and the source function closer to the Planck function and thus decreasing the line-core intensity.
One can see that the total source function at the line center sharply decreases, and that the difference between the total source function at the line center and that at the absorption trough becomes smaller. The absorption trough is still mainly formed in the photosphere, although there are some contributions from the layers around 1.1 Mm at 4 s. The formation height of the line center gradually moves higher, since the flare heating gradually enhances the contribution from the chromosphere. Because the effect of the source function decrease overweighs the effect of the opacity increase, the intensity at the line center shows a decrease with flare heating.

The second phase lasts from 4 s to 10 s. In this phase, the temperature of the chromosphere continues to rise. The ratio of $b_u/b_l$ is very close to unity, and the source function is very close to the Planck function. The contribution to the emergent intensity from the chromosphere  is now more pronounced, and the formation heights of both the line center and the absorption trough are shifted upwards. The source function in the chromosphere begins to rise, as a result of the increase of the local temperature and less deviation from LTE. Consequently, the intensities at both the line center and the absorption trough increase dramatically. In addition, the line width also increases.

\subsection{Model f11}
\begin{figure}
\epsscale{1.15}
\plotone{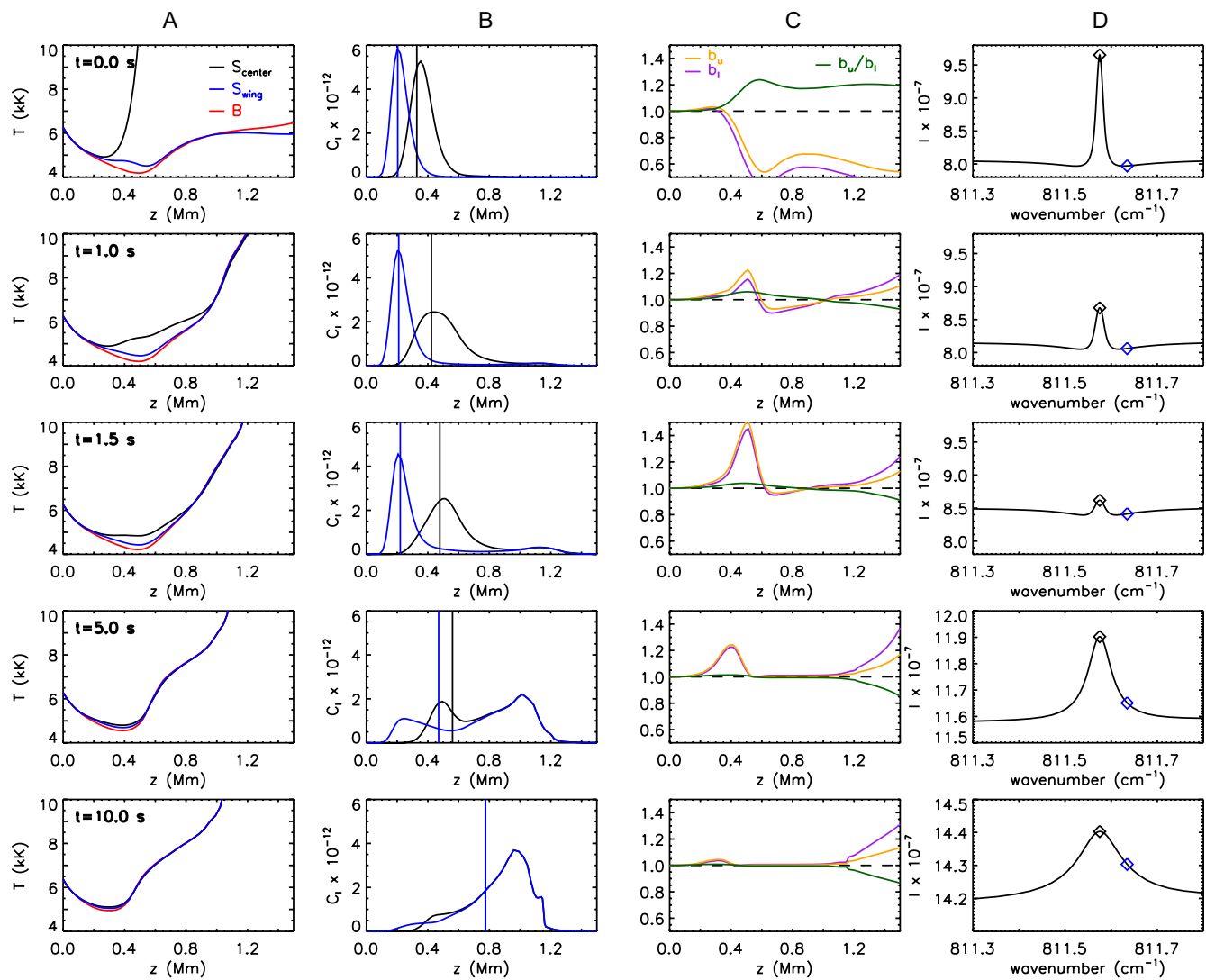}
\caption{Same as Fig.~\ref{f10}, but for the f11 model. Note that the last two panels in Column D have different intensity scales.}
\label{f11}
\end{figure}

In Fig.~\ref{f11} we show the atmospheric evolution and the line formation for the f11 model. The overall evolution pattern in this model is similar to that of the f10 model. The first phase lasts from 0.0 s to 1.5 s, when the populations at these Rydberg levels are raised, and the deviation from the LTE regime is somewhat reduced. With the ratio of $b_u/b_l$ approaching unity, the line source function gradually falls closer to the Planck function. The decrease of the total source function leads to a decrease of the intensity at the line center. During the second phase, strong heating in the chromosphere has raised the temperature dramatically. The departure coefficients of these Rydberg levels at the height of the lower chromosphere (0.6 Mm to 1.0 Mm) become very close to unity, and the source function is well coupled to the Planck function. The contribution from the lower chromosphere to the emergent intensity also increases with time, and becomes dominant at 10 s. The formation height of the line center is now shifted to the chromosphere, and the line intensity is thus greatly enhanced. The line width also increases dramatically. For example,  the wavenumber point (blue diamond in Fig.~\ref{f11}), corresponding to the absorption trough at 1.5 s, is now located at the wide emission wing at 10 s. 

\subsection{Stokes profiles}
\begin{figure}
\epsscale{1.15}
\plotone{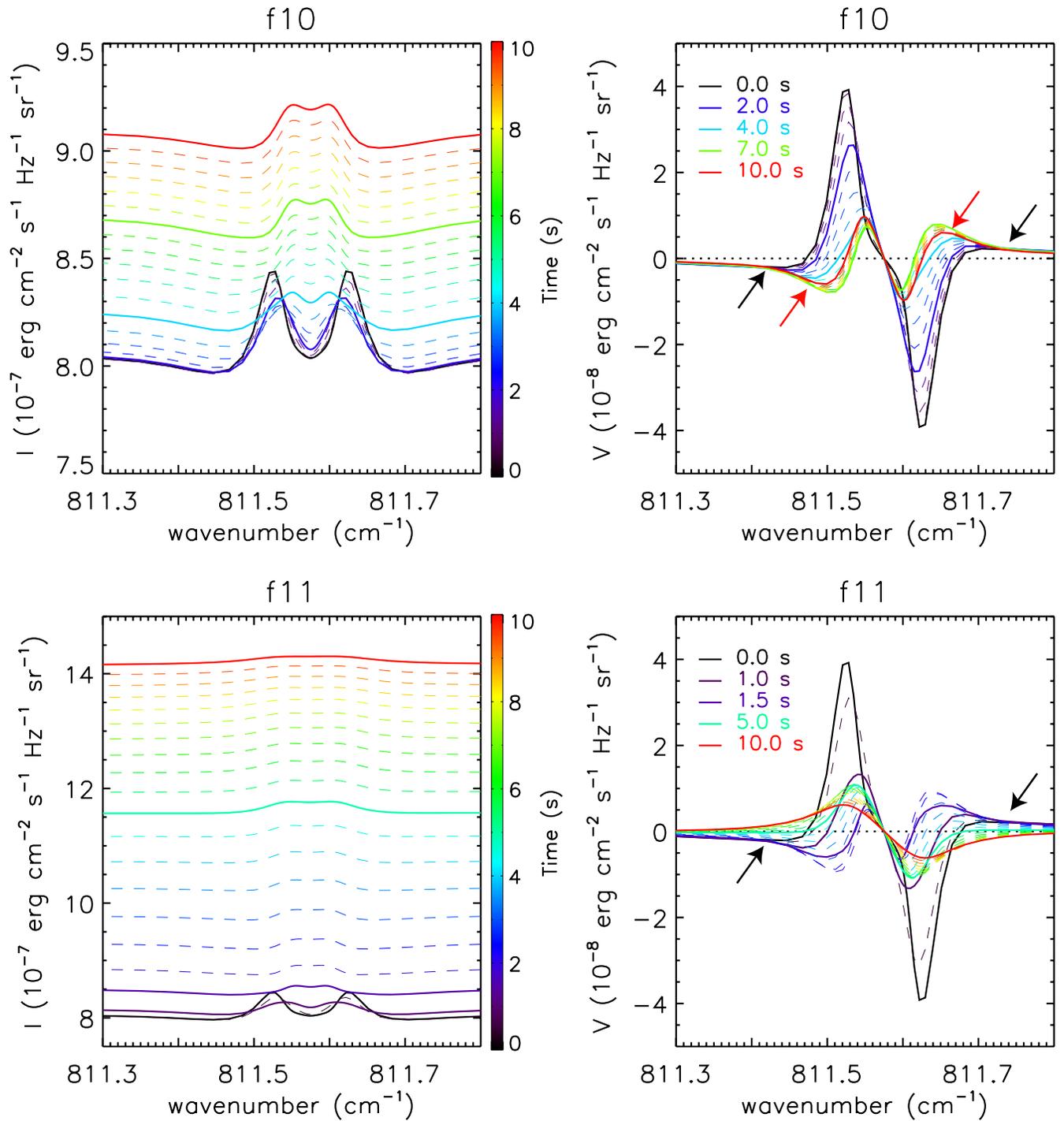}
\caption{The evolution of the Stokes $I$ and $V$ profiles, with a time step of 0.5 s. Different colors are for different time.  The solid lines correspond to the time steps chosen in Figs.~\ref{f10} and \ref{f11}. The horizontal $V=0$ line (dotted) is also plotted to guide the eye. The arrows denote the wider features.}
\label{stokes}
\end{figure}

In order to reveal the diagnostic ability of the \ion{Mg}{1} line to the magnetic field, we also calculate the Stokes profiles for this line. Fig.~\ref{stokes} shows the evolution of the Stokes $I$ and $V$ profiles. Before flare heating, the Stokes $I$ profile shows a very clear Zeeman splitting pattern with two $\sigma$ components. In the Stokes $V$ profile, apart from the two lobes that correspond to the emission features, there also exists a wide and shallow hollow in the red wing, as well as a wide and low hump in the blue wing (see the black arrows in Fig.~\ref{stokes}), just located outside the two lobes. These features are referred to as the ``wider features'' in \cite{1995bruls} and they correspond to the absorption trough in the Stokes $I$ profile. 

When heating begins, the two $\sigma$ components move closer to each other, and the intensity of the Stokes $V$ lobes decreases. As discussed above, flare heating has raised the formation height of this line, thus the observed magnetic features are less evident since the magnetic field decreases with height in our model. The Zeeman splitting pattern in the emission peaks is always visible during the whole flare evolution, although in the f11 model, the large increase in line width at a later time makes it difficult to distinguish the two $\sigma$ components. The wider features in the Stokes $V$ profile become more pronounced when heating begins. At some time the Stokes $V$ profile is very complicated with four lobes, where the wider features have a comparable intensity to the inner lobes (see the red arrows in Fig.~\ref{stokes}). The absorption trough in the Stokes $I$ profile becomes invisible at the end of the simulation period in the f11 model; correspondingly, the wider features in the Stokes $V$ profile disappear with the two lobes being increasingly broadened.

\section{Discussion}
\subsection{Eddington-Barbier relation}
The Eddington-Barbier relation states that the emergent intensity has approximately the same value as the source function at the $\tau=1$ height. This approximation becomes exact when the source function varies linearly with the optical depth. Previous studies have shown that this approximation holds for many chromospheric lines, including the H$\alpha$ line \citep{2012leenaarts,2019bjorgen}, the \ion{Mg}{2} k and h lines \citep{2013leenaarts}, and the \ion{C}{2} lines at 1334 and 1335 \AA\ \citep{2015rathore}. Here, we show the values of the emergent intensity and the source function at the height where $\tau=1$, for the \ion{Mg}{1} 12.32 $\mu$m line center in Fig.~\ref{evol}. 
It is clear that in both models these two quantities evolve oppositely at the first few seconds. The evolution of the source function begins to follow that of the intensity only at a later time. 

\subsection{Intensity dimming}
What is interesting is that at the beginning, there is a decrease of the intensity at the line center. Such an intensity dimming is also seen in the H$\alpha$ line as well as the continuum in flares and Ellerman bombs \citep{1999abbett,2005allred,2017hong,2020yang}. Generally speaking, the dimming in both the H$\alpha$ and the \ion{Mg}{1} lines are caused by the increased collisional rates during flare heating, yet the details of the dimming mechanism are not the same. For the H$\alpha$ line, the line source function is decreasing with height when it decouples with the Planck function in the mid-photosphere (at around 250 km), contributing to an absorptive line profile in the quiet Sun \citep{2012leenaarts}. Flare heating can increase the opacity through increased excitation rates, and the line center is now formed higher up and within a smaller height range \citep{2015kuridze,2019bjorgen}. The lower atmospheric layers where the source function is relatively high, now hardly have any contribution to the emergent intensity. With the upward shift of the line formation height, the local line source function is decreasing \citep{2017hong}. The Eddington-Barbier relation holds pretty well for the H$\alpha$ line, and one can instantly judge that the line intensity shall decrease. 

However, for the \ion{Mg}{1} 12.32 $\mu$m line, the line source function is increasing with height after decoupling with the Planck function in the lower photosphere (at around 100 km, see Figs.~\ref{f10} and \ref{f11}). Flare heating does increase the opacity through increased recombination and de-excitation rates, and the line formation region is slightly shifted upwards. One would definitely expect an increase in the line intensity if the source function does not change with time. However, as discussed above, the source function at the same height drops dramatically since the population departure divergence is decreased with flare heating. Thus this induces an intensity dimming since the decrease in the line source function overweighs the effect of the shift of the formation height.

Regardless of different formation mechanisms, dimming in both lines has the same response to the change of the non-thermal electron beam fluxes. When the beam flux becomes much larger, dimming is more pronounced at first, and lasts for a shorter time. In addition, dimming in the \ion{Mg}{1} 12.32 $\mu$m line has a shorter timescale than that in the H$\alpha$ line \citep{2020yang}. 

\subsection{Magnetic field inversion}
Inverting the magnetic field from the Stokes profiles is a very complicated task, while there are some simple diagnostics with certain assumptions. \cite{1995bruls} find it practical to measure the magnetic field strength directly from the Zeeman splitting in Stokes profiles. As one can easily judge from Fig.~\ref{stokes}, both the $I$ and $V$ splittings of the \ion{Mg}{1} line become smaller when heating begins. The measured magnetic field strength then decreases at an early period since the flare heating increases the formation height of the line. Such a result is similar to other lines that are used for magnetic field measurement, including the \ion{Ni}{1} 6768 \AA\ line \citep{2002ding}, and the \ion{Fe}{1} 6173 \AA\ line \citep{2018hong}.

However, after a certain time of flare heating, the line width grows, and in the f11 model the line width is so large that the absorption trough feature disappears. The $V$ splitting is now increasing, which by visual inspection gives an increasing magnetic field strength,  contrary to the fact that the line is formed higher where the magnetic field decreases. Thus we would like to note that such a simple estimation from the Zeeman splitting features might lead to erronous results, as it fails to consider the change in the line width.

We should also note the possible effect of  the shape of the line profile, especially the existence of the absorption trough, on the magnetic field measurement. We notice that in our calculations, some Stokes $V$ profiles have four lobes, which resemble the profiles that were observed and explained as originating from opposite polarities by \cite{2002jennings}. Our calculations imply that such four-lobe Stokes $V$ profiles can also be produced in the atmosphere with a single polarity, but with presence of flare-related heating. Moreover, the fact that in the line formation region, the line source function is not monotonic with regard to the optical depth, would undermine the assumption of a Milne-Eddington atmosphere. In fact, previous calculations using a Milne-Eddington atmosphere cannot reproduce the absorption trough \citep{1993hewagama}. Inversion methods with a better consideration of the atmospheric stratification \citep{2015socas,2019rodriguez} seem more promising, but new tests are still needed in the future.

\section{Conclusions}
In this paper, we calculated the \ion{Mg}{1} 12.32 $\mu$m line response in response to flare heating by electron beams using the RADYN and RH codes. Our main conclusions can be listed as follows:

1. The \ion{Mg}{1} 12.32 $\mu$m line is optically thick during the whole flare simulation. The formation height of the line center is in the upper photosphere before flare heating, but gradually shifts upwards when heating begins. At the end of the simulation time (10 s) in the strongly heated model (f11 model), the line is mainly formed in the chromosphere.

2. The population departure divergence makes the line source function decouple from the local Planck function. Initial flare heating can reduce such a decoupling to some extent, thus the source function drops close to the Planck function, resulting in a dimming at the line center.

3. The contribution to the emergent intensity from the heated layers in the chromosphere is increasing with time. At the end of the simulation time of the f11 model, the chromospheric contribution is dominant, and the line width also increases dramatically. 

4. The change in the line formation height during flare heating can significantly influence the Stokes profiles. The Zeeman splitting width in the Stokes $I$ profile becomes less pronounced and the lobe intensity in the Stokes $V$ profile decreases when flare heating sets in.

5. The Stokes profiles can be very complicated during a flare. The existence of the absorption trough (the wider features in the Stokes $V$ profile) makes it difficult to invert the magnetic field using a simple Milne-Eddington model.

We conclude that this line is very sensitive to flare heating, and can present very complicated Stokes profiles. Thus when measuring the magnetic field in a flare with this line, one needs to be very careful to distinguish the real signals related to the magnetic field from the artifacts induced by flare heating. As such, a more advanced model that can include the complexity of the source function is required in order to properly invert the magnetic field.

\acknowledgments
We thank the referee for detailed suggestions that help improve the paper. J.H. would like to thank Yang Guo for helpful discussions. This work was supported by NSFC under grants 11903020, 11733003, 11427901, 11873062, 11873095, 11533005, and 11961131002, and NKBRSF under grant 2014CB744203. J.H. is also supported by CAS Key Laboratory of Solar Activity under grant KLSA201904. Y.L. is supported by the CAS Pioneer Talents Program for Young Scientists and XDA15052200, XDA15320301, and XDA15320103.

\clearpage

\end{document}